%% file: main_v4.tex
\documentclass[%
 reprint,
superscriptaddress,
bibnotes,
 amsmath,amssymb,
 aps,
]{revtex4-2}

\usepackage{graphicx}
\usepackage{dcolumn}
\usepackage{bm}
\usepackage{makecell}
\usepackage{xcolor}
\usepackage{multirow}
\usepackage{physics}
\usepackage{tikzorbital}

\usepackage{float}
\input{box_1/boxmacros}
\renewcommand{\boxcaptiontext}{Equations defining the representative physical phenomena. THG = third harmonic generation; EFISH = electric-field induced second harmonic; (D)FWM = (degenerate) four-wave mixing.}
\renewcommand{\boxlabelname}{box:definitions}

\begin{document}

\date{\today}
\title{Universal unconventional responses controlled by ferroaxial order}

\author{Zhiren He}
 \email{zhiren.he@rug.nl}
\affiliation{Zernike Institute for Advanced Materials, University of Groningen, Nijenborgh 3, 9747 AG Groningen, The Netherlands}
\affiliation{Department of Physics, University of North Texas, Denton, TX 76203, USA}

\author{Guru Khalsa}%
\affiliation{Department of Physics, University of North Texas, Denton, TX 76203, USA}

\author{Jagoda S{\l}awi{\'n}ska}
\affiliation{Zernike Institute for Advanced Materials, University of Groningen, Nijenborgh 3, 9747 AG Groningen, The Netherlands}

\begin{abstract}

Ferroaxial materials are an emerging class of ferroic materials that, in contrast to ferromagnets and ferroelectrics, are fully robust against stray fields, making them ideally suited for data storage and other nonvolatile applications. However, detection of the ferroaxial state remains challenging, as ferroaxiality does not manifest directly through a measurable macroscopic electric polarization or magnetization, and probes are mostly limited to optical methods. Here, we theoretically establish that ferroaxial order universally generates unconventional components of responses to external stimuli, such as the spin Hall effect, magneto-Seebeck effect, or Faraday effect, across a broad range of linear and nonlinear transport, optical and equilibrium phenomena. These unconventional responses are always directly coupled to ferroaxial order and can distinguish ferroaxial domains, providing probes of ferroaxial order. This general connection also reveals a largely unexplored class of ferroaxial metals, in which unconventional transport provides a natural probe of ferroaxial order, as demonstrated by our first-principles calculations for representative materials. Our results reveal a fundamentally different way for ferroic order to manifest -- ferroaxiality primarily shows unconventional material responses rather than the directly measurable order parameters, broadening the possibilities of nonvolatile ferroic control and flexible device design.
\end{abstract}
\maketitle

\begin{figure*}[htbp]
    \centering
    \includegraphics[width=\textwidth]{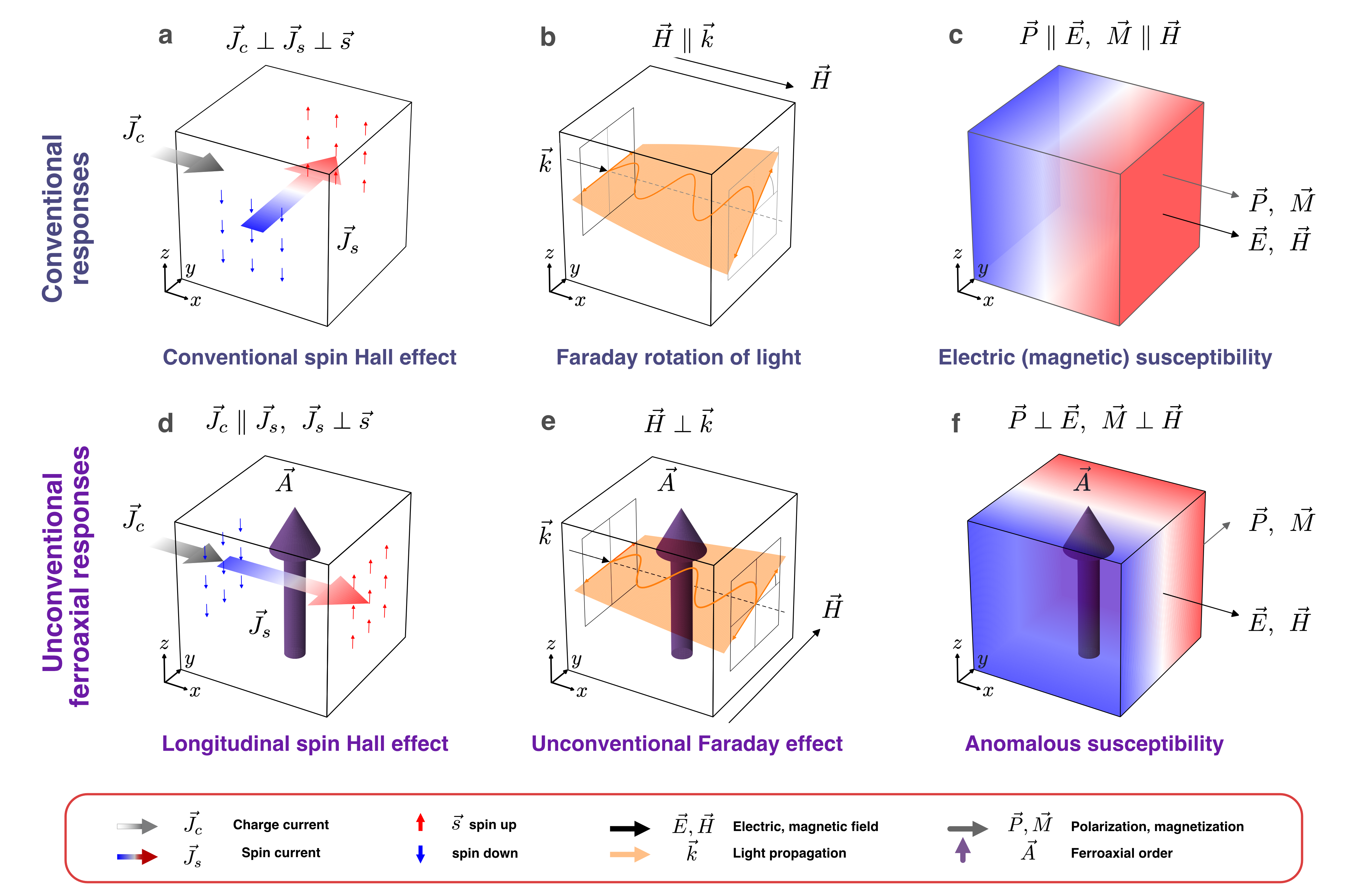}
    \caption{\textbf{Conventional and unconventional responses to external stimuli.} (a) Conventional spin Hall effect (SHE) where charge current, spin current, and their spin polarization are mutually perpendicular. (b) Faraday rotation of light propagating along the direction of the magnetic field. (c) Induced electric polarization (magnetization) collinear with the electric (magnetic) field. (d) Unconventional SHE with parallel charge and spin current. (e) Unconventional Faraday effect where light perpendicular to the magnetic field acquires polarization rotation. (f) Anomalous transverse susceptibility leading to a perpendicular component of induced electric polarization (magnetization). In the bottom panels, the ferroaxial order $\vec{A}$ unlocks unconventional components of response, in addition to conventional ones.}
    \label{fig:unconventional}
\end{figure*}

Ferroic orders in solids enable collective switching that is a cornerstone of modern technologies, including memories and emerging computing devices. Ferroaxiality has been recently identified as a distinct ferroic order, characterized by a rotational head-to-tail arrangement of electric dipoles that preserves both spatial inversion ($\mathcal{I}$) and time reversal ($\mathcal{T}$) symmetry \cite{Dubovik1986, Hlinka_2014, Hlinka_2016}. In contrast to ferroelectrics and ferromagnets, ferroaxial materials lack net polarization, which makes them insensitive to electric and magnetic fields, resulting in the absence of stray and depolarizing fields. The recent discovery of artificial conjugate fields and the experimental demonstration of photo-induced switching of ferroaxial domains have established ferroaxiality as a controllable ferroic order and opened the way for practical device realizations \cite{He_2024, Zeng_2025}.

Response tensors provide the mathematical description of how crystals respond to external stimuli. The symmetry of the material is therefore reflected in the physical responses it can exhibit. While ferroaxial materials may appear inert at first glance owing to their preserved inversion and time reversal symmetries, the distinctive spatial distribution of charges gives rise to phenomena that are unusual in high-symmetry phases, providing probes of the ferroaxial state. For example, ferroaxial order reshapes the electric quadrupole second-harmonic generation (SHG) and linear electrogyration tensors, enabling optical imaging of ferroaxial domains \cite{Jin_2019,Yokota_2022,Guo_2023}. More recently, an anomalous transverse susceptibility (ATS) was identified as a surprising electromagnetic consequence of ferroaxial order \cite{Inda_2023,Du_2025}. For practical devices, however, electrical detection of the ferroaxial state through transport would be desirable. Analyses of model systems have revealed unconventional Hall and spin Hall effects \cite{Hayami_2022,Hayami_2023}, and piezoresistivity has been proposed as a fingerprint of the ferroaxial transition \cite{Day_Roberts_2025}. Yet the ferroaxial materials identified to date are insulating, leaving such responses largely unexplored in realistic materials.

Here, we use group theory to establish a universal connection between ferroaxial order and unconventional responses to external stimuli across a broad range of condensed matter phenomena, including electrical, thermal, spin, optical, magnetic, and electromechanical effects. Figure 1 illustrates this concept for representative transport, optical, and equilibrium phenomena by comparing conventional responses with their unconventional ferroaxial counterparts. The unconventional response tensor components generated by ferroaxial order are directly coupled to the ferroaxial state and always reverse sign for opposite domains, enabling nonvolatile control of the responses. This connection also expands the range of physical phenomena that can serve as probes of ferroaxial order. We then examine the spin Hall conductivity (SHC) as a representative rank-3 pseudotensor and use a tight-binding model to reveal the microscopic origin of its ferroaxial components. Finally, we identify several families of ferroaxial metals and perform first-principles calculations for representative materials showing that unconventional spin Hall conductivity and piezoresistivity can distinguish between opposite ferroaxial domains.

\section*{Results}
\subsection{Symmetry relation between ferroaxial order and unconventional response}
Box \ref{box:definitions} summarizes response tensors corresponding to representative condensed matter phenomena. Our analysis presented below is applicable to all of them and also to other phenomena not included in the list. The ferroaxial components of response tensors are identified in three steps. We first establish the most general connection between response tensors and ferroaxial order using representation theory, without restricting to any particular crystal symmetry, and show that response tensors contain unconventional components coupled to ferroaxial order. We then impose the intrinsic symmetries associated with the specific physical response, which can arise from conditions such as microscopic reversibility and can eliminate these ferroaxial components in some cases. Finally, we consider the crystallographic point group symmetries, where ferroaxial components can be recovered even when they are absent at the level of the full rotation group. 
\vskip 0.4em
\textbf{Tensor representation of ferroaxial order}. We formulate the connection between ferroaxial order and response tensors in a general way, without restricting the analysis to a particular physical effect. We focus on $\mathcal{T}$-even and $\mathcal{I}$-even tensors because ferroaxiality, behaving like an axial (pseudo) vector, is invariant under inversion and time reversal symmetries. Although ferroaxial order may also manifest in $\mathcal{T}$- and/or $\mathcal{I}$-odd tensors, such contributions require crystals with broken $\mathcal{T}$ and/or $\mathcal{I}$ symmetry and are not considered here.

\input{box_1/box}

Two simple definitions of ferroaxial order $\bm{A}$ are given by the cross product of two polar vectors: position vector and electric dipole moment $\bm{A}=\bm{r}\times \bm{d}$, and the cross product of two axial vectors: orbital and spin angular momentum $\bm{A}=\bm{\ell}\times \bm{s}$ \cite{Hayami_2018,Kusunose_2020}. These cross products correspond to the antisymmetric part of a rank-2 tensor and thus provide the simplest tensor representation of ferroaxial order. In the language of group theory, an $\mathcal{I}$-even rank-2 tensor $B$, described by the Jahn symbol $V^2$, is constructed from the outer product of either two polar vectors or two axial vectors:
\begin{equation}
B=V^-\otimes V^-=V^+\otimes V^+,
\end{equation}
where we use $V^-$($V^+$) to denote the irreducible representation (irrep) of a polar (axial) vector, odd (even) under inversion. The Clebsch-Gordan decomposition of $B$ is
\begin{equation}
    B=\Gamma_{\ell=2}^+\oplus\Gamma_{\ell=1}^+\oplus\Gamma_{\ell=0}^+, \label{eq:V2}
\end{equation}
consisting of a symmetric part $\Gamma_{\ell=2}^+\oplus\Gamma_{\ell=0}^+$ and an antisymmetric part $\Gamma_{\ell=1}^+$, with $\Gamma$ being the irrep of the full rotation group SO(3). Since the antisymmetric part $\Gamma_{\ell=1}^+$ transforms as an axial vector $V^+$, it shares the same irrep as ferroaxial order. The components of ferroaxial order $A_i$ ($i=x,y,z)$ are therefore linked to the corresponding antisymmetric tensor components, which form the axial vector $\eta_i=\varepsilon_{ijk}B_{jk}$ ($\varepsilon_{ijk}$ is the Levi-Civita symbol).

In Landau theory, this link allows a bilinear coupling in free energy $\mathcal{F}\propto \bm{A}\cdot \bm{\eta}$, between the ferroaxial order $\bm{A}$ and the corresponding components of the response tensor $\bm{\eta}$. Thus, a change in $A_x$ gives rise to a change in the response $B_{yz}=-B_{zy}$, with analogous relations holding for $A_y$ and $A_z$.

A general $\mathcal{I}$-symmetric response tensor of rank $n$ without intrinsic permutation symmetry can be written as
$(V^+)^{\otimes n}$, containing at least one component $\bm{\eta}$ that transforms as the axial vector $V^+$. Therefore, any general response tensor of rank $\geq 2$ contains components that can completely capture the ferroaxial order and distinguish its order-parameter direction ($x$, $y$, $z$). In other words, every general response tensor necessarily contains components that are controlled by ferroaxial order.

\begin{table*}[hbt!]
    \centering 
    \begin{tabular*}{\textwidth}{@{\extracolsep{\fill}}cccccc} \hline \hline
        Rank & Jahn symbol & Contain $V^+$ & Transport tensor & Optical tensor & Equilibrium tensor\\ \hline 
        2 & $[V^2]$  & N & Electric resistivity  & --  & 
        Dielectric and magnetic susceptibility
        \\ \cline{2-6} 
        & $V^2$ & Y & Seebeck effect & -- & -- \\
        & & & Peltier effect \\ \hline
        3 & $eV^3$ &  Y & Spin Hall effect & $\chi_2^m$ & --\\ 
        & & & Nernst effect \\
        & & & Ettinghausen effect \\\cline{2-6} 
         & $e[V^3]$ &  Y & -- & $\chi_2^m$ (Kleinman) & --\\ \cline{2-6}
         & $e[V^2]V$ &  Y & -- & Electrogyration effect & -- \\ \cline{2-6} 
         & $e\{V^2\}V$ & Y & Hall effect & Faraday effect & -- \\ 
         & & & Thermal Hall effect \\ \hline
        4 & $V^4$ & Y &$-i\omega\chi_3$ & $\chi_3$ & --\\ \cline{2-6} 
         & $[V^4]$ & N & -- & $\chi_3$ (Kleinman) & 
         Third-order dielectric susceptibility 
         \\ 
         & & & & THG (Kleinman) & Third-order magnetic susceptibility \\
         & & & & EFISH (Kleinman) \\ \cline{2-6} 
         & $[V^2][V^2]$ & Y & Piezoresistivity  & Kerr effect & Electrostriction\\ 
         & & & Magnetic resistance & Cotton-Mouton effect & Magnetostriction \\
         & & & & Elasto-optical effect \\ \cline{2-6}
         & $V^2[V^2]$ & Y & $-i\omega\chi_3^{EFISH}$ & $\chi_3^{EFISH}$ & Flexoelectricity \\ 
         & & & Magneto-Seebeck effect \\
         & & & Magneto-Peltier effect\\ \cline{2-6} 
         & $[[V^2][V^2]]$ & N & $-i\omega\chi_3^{DFWM}$ & $\chi_3^{DFWM}$ & Elastic compliance\\ \cline{2-6} 
         & $[V^2V^2]$ & Y & $-i\omega\chi_3^{FWM}$ & $\chi_3^{FWM}$ & -- \\ \cline{2-6} 
         & $V[V^3]$ & Y & $-i\omega\chi_3^{THG}$ & $\chi_3^{THG}$ & -- \\ \hline \hline
    \end{tabular*}
    \caption{List of $\mathcal{T}$- and $\mathcal{I}$-symmetric tensors up to rank 4, compiled with the information from TENSOR routine under Bilbao Crystallographic Server \cite{Gallego_2019}. Square/curly bracket ([]/\{\}) means the tensor is symmetric/antisymmetric under the permutation of the indices. The third column shows whether the representation of the tensor in the full rotation group contains at least one axial vector irrep $V^+=\Gamma_{\ell=1}^+$. $\chi^m_2$ denotes second-order susceptibility by magnetic dipole. Several $\chi_3$ tensors represent various third-order susceptibility in nonlinear optics, with $-i\omega\chi_3$ tensors the corresponding counterpart in nonlinear transport.}
    \label{tab:effects}
\end{table*}
\vskip 0.4em
\textbf{Intrinsic tensor symmetries.} The form of a response tensor can be further constrained by the fundamental physical principles behind the response. For example, Onsager reciprocity in transport imposes specific symmetry relations between response coefficients. In nonlinear optics, the responses can likewise exhibit Kleinman symmetry, which makes the tensor invariant under permutations of its indices when the relevant frequencies are sufficiently far from resonances and absorption can be neglected \cite{Kleinman_1962}. As a result, these intrinsic symmetries reduce the representation space, and the $V^+$ component identified above may either be retained or eliminated. Applying intrinsic constraints for the response tensors considered in Box \ref{box:definitions}, we identify which physical phenomena retain $V^+$ in the full rotation group, and consequently in all point groups, as summarized in Table \ref{tab:effects}. Thus, these phenomena are able to \emph{universally} host unconventional components of ferroaxial origin. 

A notable case are rank-2 tensors. For equilibrium tensors such as static dielectric and magnetic susceptibility $\chi^{e,eq}_{ij}$ and $\chi^{m,eq}_{ij}$, the symmetric condition follows from the equality of mixed derivatives of the free energy $\mathcal{F}$, as \[\chi^{e,eq}_{ij}=\frac{\partial P_i}{\partial E_j}=-\frac{\partial^2\mathcal{F}}{\partial E_i\partial E_j},\quad\chi^{m,eq}_{ij}=\frac{\partial M_i}{\partial H_j}=-\frac{\partial^2\mathcal{F}}{\partial H_i\partial H_j}.\] At finite frequency or in transport, for example, electric resistivity $\rho_{ij}$, the symmetric condition follows instead from the Onsager reciprocity relation: $L_{ij}(\omega,H)=L_{ji}(\omega,-H)$ for any dynamical linear-response coefficient $L$ \cite{Onsager_1931, Casimir_1945}. As a result of the intrinsic symmetry that forbids the antisymmetric part, symmetric rank-2 tensors ($[V^2]$) lose $V^+$ (Eq. \ref{eq:V2}) and therefore do not universally reflect ferroaxial order in all point groups. 

A distinct case is the pair of thermoelectric tensors, the Seebeck coefficient $\beta_{ij}$ and Peltier coefficient $\pi_{ij}$, related by $\beta_{ij}=\pi_{ji}$ in nonmagnetic systems \cite{Gallego_2019}. Although the antisymmetric part is allowed, it is usually only manifested in monoclinic and triclinic point groups, where the principal axes of these tensors do not necessarily align \cite{Kaganov_1979}.
\vskip 0.4em
\textbf{Tensor decomposition.} Here we explicitly work out the representation of rank-3 pseudo-tensors with intrinsic symmetries. A general rank-3 pseudo-tensor expressed as: \begin{equation}
    eV^3=\Gamma_{\ell=3}^+\oplus 2\Gamma_{\ell=2}^+\oplus 3\Gamma_{\ell=1}^+\oplus\Gamma_{\ell=0}^+ \label{eq:eV3}
\end{equation} can be split into the symmetric part \begin{equation}
    e[V^2]V=(\Gamma_{\ell=2}^+\oplus\Gamma_{\ell=0}^+)\otimes\Gamma_{\ell=1}^+=\Gamma_{\ell=3}^+\oplus \Gamma_{\ell=2}^+\oplus 2\Gamma_{\ell=1}^+
\end{equation} and antisymmetric part \begin{equation}
    e\{V^2\}V=\Gamma_{\ell=1}^+\otimes\Gamma_{\ell=1}^+=\Gamma_{\ell=2}^+\oplus \Gamma_{\ell=1}^+\oplus \Gamma_{\ell=0}^+
\end{equation} with respect to two of the indices. This demonstrates the presence of $V^+$ in both symmetric and antisymmetric parts, and suggests a variety of rank-3 responses, as listed in Table I, including the spin Hall effect, and the Hall effect or Faraday effect, whose unconventional components can serve as a probe of ferroaxial order. 

Repeating similar analysis for rank-4 tensors -- see Supplementary Material (SM), Sec. I -- reveals that this universality also holds for many of the rank-4 responses (``Y" in Table \ref{tab:effects}). However, as our analysis is based on the full rotation group, the presence of $V^+$ is a sufficient but not necessary condition for the tensor to contain a ferroaxial irrep upon symmetry lowering to a crystallographic point group, i.e. the absence of $V^+$ under full rotation group (``N" in Table \ref{tab:effects}), does not rule out ferroaxial irreps in crystallographic point groups. This is because distinct irreps of the full rotation group decompose into sets of point group irreps that may sometimes overlap. Consequently, a point group irrep associated with ferroaxial order can arise from full-rotation-group irreps other than $V^+$. For instance, the equilibrium third-order susceptibility tensor $\chi^{e,eq}_3$ and $\chi^{m,eq}_3$ (Jahn symbol [$V^4$]) lacks $V^+$ in the full rotation group. However, in tetragonal crystals, a ferroaxial order along the 4-fold rotation axis $C_{4z}$ will induce the third-order anomalous transverse component of dielectric and magnetic susceptibility $\chi^{e,eq}_{yxxx}$ and $\chi^{m,eq}_{yxxx}$ illustrated in Fig. \ref{fig:unconventional}f \cite{Inda_2023, Du_2025}, as discussed below. 

\vskip 0.4em
\textbf{Example of point group 4/mmm}. Taking the point group 4/mmm as an example, the three-dimensional axial-vector representation $\Gamma_{\ell=1}^+$ is decomposed into $E_g\oplus A_{2g}$. A general rank-3 pseudo-tensor $\chi$ can then be expressed in terms of 4/mmm irreps as
$(E_g\oplus A_{2g})^{\otimes 3}=3A_{1g}\oplus 4A_{2g}\oplus 7E_g\oplus 3B_{1g}\oplus 3B_{2g}$. As expected, both the symmetric part $e[V^2]V=A_{1g}\oplus 3A_{2g}\oplus 5E_g\oplus 2B_{1g}\oplus 2B_{2g}$ and the antisymmetric part $e\{V^2\}V=2A_{1g}\oplus A_{2g}\oplus 2E_g\oplus B_{1g}\oplus B_{2g}$ contain $E_g$ and $A_{2g}$. Note that although Eq. \ref{eq:eV3} only contains 3 copies of $\Gamma_{\ell=1}^+$, additional $A_{2g}$ and $E_g$ come from $\Gamma_{\ell=3}^+=A_{2g}\oplus2E_g\oplus B_{1g}\oplus B_{2g}$, as well as $\Gamma_{\ell=2}^+=A_{1g}\oplus E_g\oplus B_{1g}\oplus B_{2g}$. This shows that even if $V^+$ is absent in the tensor representation under the full rotation group, a ferroaxial irrep may still show up in certain point groups.

Once the relevant irreps have been identified, the tensor components associated with each irrep can be found using the projection operator \cite{Dresselhaus_2008,ISOTROPY}. Since the tensor representation consists of 3 copies of the identity irrep $A_{1g}$, one in $e[V^2]V$ and two in $e\{V^2\}V$, a few tensor elements can already be nonzero in point group $4/mmm$:
\begin{eqnarray*}
\chi_1&=&\chi^z_{xy}=-\chi^z_{yx} \\
\chi_2&=&\chi^y_{zx}=-\chi^y_{xz}=\chi^x_{yz}=-\chi^x_{zy}\\
\chi_3&=&\chi^y_{zx}=\chi^y_{xz}=\chi^x_{yz}=\chi^x_{zy}.
\end{eqnarray*}
Here, indices with permutation symmetry are in the subscript, distinguishing them from the third index in the superscript. With the appearance of ferroaxial order $\bm{A}=(0,0,A_z)$ that lowers the symmetry, 4 independent new components that transform as $A_{2g}$ emerge:
\begin{eqnarray*}
A_{1z} &=& \chi^z_{xx}=\chi^z_{yy} \\
A_{2z} &=& \chi^z_{zz} \\
A_{3z} &=& \chi^y_{yz}=\chi^y_{zy}=\chi^x_{xz}=\chi^x_{zx} \\
A_{4z} &=& \chi^y_{yz}=-\chi^y_{zy}=\chi^x_{xz}=-\chi^x_{zx}
\end{eqnarray*}
where $A_{1z},A_{2z},A_{3z}$ represent the symmetric part, and $A_{4z}$ represents the antisymmetric part, consistent with the number of $A_{2g}$ irreps in the representation. Thus, the form of the tensor in the presence of a ferroaxial order $A_z$ is:
\[\left(\begin{array}{c|cccccc|ccc} & xx & yy & zz & [yz] & [zx] & [xy] & \{yz\} & \{zx\} & \{xy\} \\ \hline
x & 0 & 0 & 0 & \chi_3 & A_{3z} & 0 & \chi_2 & -A_{4z} & 0\\
y & 0 & 0 & 0 & A_{3z} & \chi_3 & 0 & A_{4z} & \chi_2 & 0 \\
z & A_{1z} & A_{1z} & A_{2z} & 0 & 0 & 0 & 0 & 0 & \chi_1
\end{array}\right),\]
revealing how the ferroaxial order is reflected by each independent component of a response tensor. 

Consider the Hall effect $E_i=R_{ij}^kJ_jH_k$ as an example. The tensor $R_{ij}^k$ is antisymmetric under the exchange of $i$ and $j$, with Jahn symbol $e\{V^2\}V$. A ferroaxial order along $z$ direction will activate the component $A_{4z}$, i.e. 4 tensor elements are activated with the relation $R^y_{yz}=-R^y_{zy}=R^x_{xz}=-R^x_{zx}$. In SM (Sec. II), we extend the explicit analysis to all ferroaxial point groups for rank-3 tensors with intrinsic symmetry. 

\subsection{Ferroaxial effective Hamiltonian}
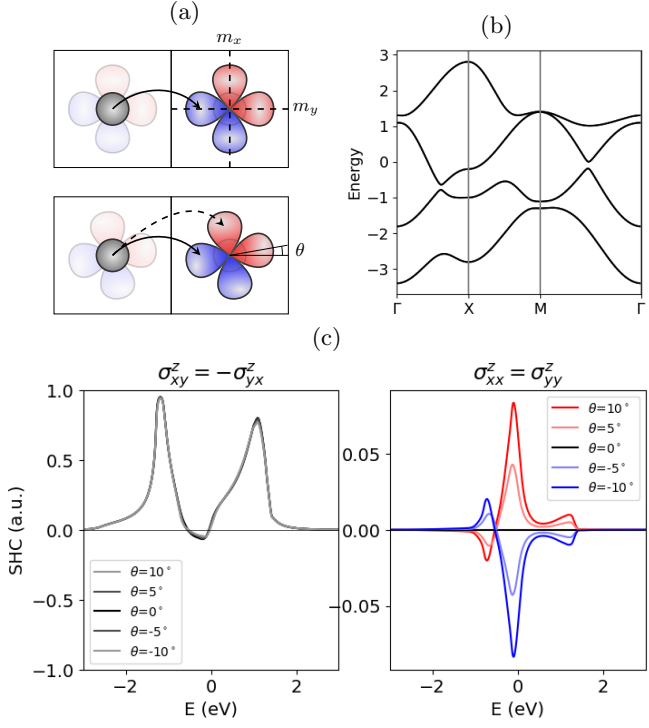
\begin{figure}[htb!]
\centering 
\hfill
\begin{minipage}[b]{0.45\linewidth}
    (a)
    \centering
    \resizebox{\linewidth}{!}{
    \begin{tikzpicture}[scale=1]
	\draw (0,3) rectangle (2,5);
	\draw (2,3) rectangle (4,5);
	\orbital[pos = {(1,4)}, color=gray]{s}
	\orbital[pos = {(3,4)}, color=gray, opacity=0.2]{s}
	\orbital[pos = {(3,4)}, pcolor=red, ncolor=blue]{py}
	\orbital[pos = {(3,4)}, pcolor=red, ncolor=blue]{pz}
	\orbital[pos = {(1,4)}, pcolor=red, ncolor=blue, opacity=0.2]{py}
	\orbital[pos = {(1,4)}, pcolor=red, ncolor=blue, opacity=0.2]{pz}
	\draw[dashed,thick] (2,4) -- (4,4);
	\draw[dashed,thick] (3,3) -- (3,5);
	\node[centered] at (4.3,4) {$m_y$};
	\node[centered] at (3,5.2) {$m_x$};
	
	\draw (0,0.5) rectangle (2,2.5);
	\draw (2,0.5) rectangle (4,2.5);
	\orbital[pos = {(1,1.5)}, color=gray]{s}
	\orbital[pos = {(3,1.5)}, color=gray, opacity=0.2]{s}
	\begin{scope}[shift={(3,1.5)},rotate=10]
		\orbital[pos = {(0,0)}, pcolor=red, ncolor=blue]{py}
		\orbital[pos = {(0,0)}, pcolor=red, ncolor=blue]{pz}
	\end{scope}
	\begin{scope}[shift={(1,1.5)},rotate=10]
		\orbital[pos = {(0,0)}, pcolor=red, ncolor=blue, opacity=0.2]{py}
		\orbital[pos = {(0,0)}, pcolor=red, ncolor=blue, opacity=0.2]{pz}
	\end{scope}

	\draw[-stealth, thick]
	(1,4) to[out=45,in=135] (2.5,4);   
	\draw[-stealth, thick, dashed]
	(1,1.5) to[out=45,in=135] (2.9,2); 
	\draw[-stealth, thick]
	(1,1.5) to[out=45,in=135] (2.5,1.5); 
	\draw (3.9, 1.5) arc (0:10.0:0.9);
    \draw (3,1.5) -- (4,1.5);
    \draw (3,1.5) -- (3.9848,1.6736);
    \node[centered] at (4.2,1.6) {$\theta$};
    \end{tikzpicture}
    }
\end{minipage}
\begin{minipage}[b]{0.48\linewidth}
    (b)
    \centering
    \includegraphics[width=\linewidth]{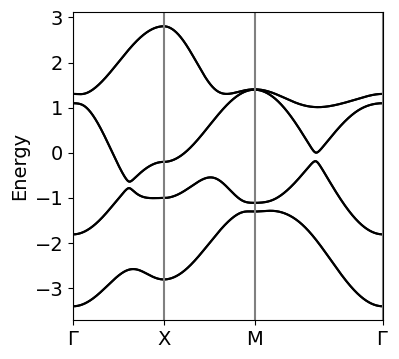}
\end{minipage} \vspace{0.5em}
\begin{minipage}{\linewidth}
    (c)
    \centering
    \includegraphics[width=\linewidth]{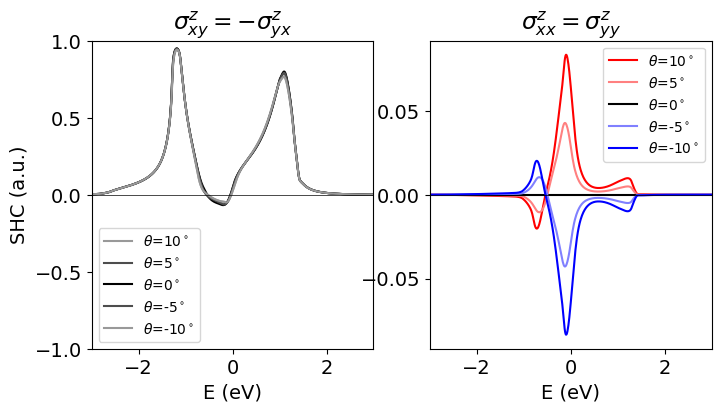}
\end{minipage}
\caption{\textbf{Numerical study based on a 2D effective model.} (a) Schematic image showing hopping between $s$ and $p_x/p_y$ orbitals in para-axial phase and additional hopping with the onset of ferroaxial order $A_z$ that rotates the $p_x,p_y$ orbitals by angle $\theta$ and breaks all the vertical mirror symmetries. (b) Band structure derived from the tight-binding model in the para-axial phase, i.e., when $\theta=0$.  (c) Spin Hall conductivity (in arbitrary units) with different values of rotation angle $\theta$. Model parameters are defined in SM (Sec. V). \label{fig:TB}}
\end{figure}

To gain insight into how changes in bonding give rise to unconventional components of response tensors, we use the method of invariants \cite{Bir_1974,Linnik_2026} to construct a minimal model of $s$ and $p$ orbitals with spin (dimension = $8\times8$) on a tetragonal lattice. We study a ferroaxial transition from point group 4/mmm to 4/m with the order parameter $A_z$ oriented along the $z$ axis (irrep $A_{2g}$) and determined by the rotation angle $\theta$ (Fig. \ref{fig:TB}a). A two-dimensional (2D) model that captures the essential physics is presented here, and the complete symmetry analysis and numerical results for a three-dimensional (3D) model are provided in the SM (Sec. III and IV).

Defining orbital couplings $T_\alpha \equiv (\ket{s}\bra{p_\alpha},\ket{p_\alpha}\bra{s}),~\alpha=\{x,y,z\}$ and $L_\gamma \equiv i(\ket{p_\alpha}\bra{p_\beta}-\ket{p_\beta}\bra{p_\alpha}),~\alpha\neq\beta\neq\gamma$, the general form of the terms that appear in the ferroaxial phase to the lowest order are $i\bm{A}\cdot(\bm{k}\times\bm{T})$, and $i\bm{A}\cdot(\bm{L}\times\bm{\sigma})$ --
the cross product of two polar vectors and two axial vectors, matching the two definitions of ferroaxial order. When projected onto the $z$-axis, they give $iA_z(k_xT_y-k_yT_x)$ and $iA_z(L_x\sigma_y-L_y\sigma_x)$. One comes from new bonding channels that form when rotation of $p$ orbitals breaks the vertical mirror symmetries (dashed arrow in Fig. \ref{fig:TB}a) and the other from the change in local crystal field potential $\nabla V$. We note that vectorial operators $\bm{T}$ and $\bm{L}$ are important, as otherwise the lowest order term in a tetragonal lattice that appears in the ferroaxial phase would be the hexadecapolar $\ell=4$ terms $(\ket{p_x}\bra{p_x}-\ket{p_y}\bra{p_y})k_xk_y$ and $(\ket{p_x}\bra{p_y}+\ket{p_y}\bra{p_x})(k_x^2-k_y^2)$, resembling the $d_{x^2-y^2}$-$d_{xy}$ mixing in the $d$-orbital model via the ferroaxial crystal field \cite{Hayami_2022,Hayami_2023}. This shows that hybridization between orbitals of opposite parity ($s-p$, $p-d$, etc.) is an important microscopic feature of the ferroaxial transition, consistent with previous studies of ferroaxial systems \cite{Bhowal_2024,Inda_2025}. 

In order to generate a spin Hall response, the velocity operator, $v_\alpha=\partial\mathcal{H}/\partial k_\alpha$, must acquire a spin-dependent contribution. Neither the intercell $s-p$ hopping term $i\bm{A}\cdot(\bm{k}\times\bm{T})$ nor the antisymmetric spin-orbit coupling (SOC) term $i\bm{A}\cdot(\bm{L}\times\bm{\sigma})$ alone satisfies this requirement, as the former is spin independent, and the latter carries no $\bm{k}$ dependence. Nevertheless, both can contribute indirectly to the SHC through their interplay with other terms in the Hamiltonian. For example, a term $(k_xT_x+k_yT_y)\sigma_z$ can be derived from $i(k_xT_y-k_yT_x)$ in combination with SOC $L_z\sigma_z$, contributing to the unconventional SHC component $\sigma^z_{xx}=\sigma^z_{yy}$. Likewise, the combination of antisymmetric SOC with $i\bm{k}\cdot\bm{T}$ generates spin-dependent hopping that contribute to unconventional $\sigma^{y}_{yz}=\sigma^{x}_{xz}$ and $\sigma^{y}_{zy}=\sigma^{x}_{zx}$ components, whose description requires the full 3D Hamiltonian. A detailed analysis is provided in SM (Sec. IV). These spin-dependent terms also emerge naturally in our full effective Hamiltonian, together with other additional symmetry-allowed terms, as shown in Table S5 and S6.

Next, to verify our analysis, the 2D effective Hamiltonian with one set of chosen parameters is converted to a tight-binding model (see SM, Sec. V) and constructed using PythTB \cite{PythTB}. To obtain the SHC from this model, we develop an interface between PythTB and PAOFLOW \cite{Buongiorno_Nardelli_2018,Cerasoli_2021}. The resulting band structure is shown in Fig. \ref{fig:TB}b and the SHC in Fig. \ref{fig:TB}c. In the band structure, the states are always doubly degenerate because both $\mathcal{T}$ and $\mathcal{I}$ are present. For a 2D model, only two independent components $\sigma^z_{xy}=-\sigma^z_{yx}$ and $\sigma^z_{xx}=\sigma^z_{yy}$ are allowed in the ferroaxial phase. Thus, when $\theta=0$, all symmetries of 4/mmm are preserved and only the conventional components $\sigma^z_{xy}=-\sigma^z_{yx}$ survive. As we vary the rotation angle, there is little change in the band structure and the conventional components, but the unconventional components $\sigma^z_{xx}=\sigma^z_{yy}$ appear. As expected, the sign and magnitude of the ferroaxial order $A_z\sim\sin\theta$ control the sign and magnitude of the unconventional components.

\subsection{First-principles studies of ferroaxial metals}

\begin{figure*}[!htb]
    \centering
    \includegraphics[width=\textwidth]{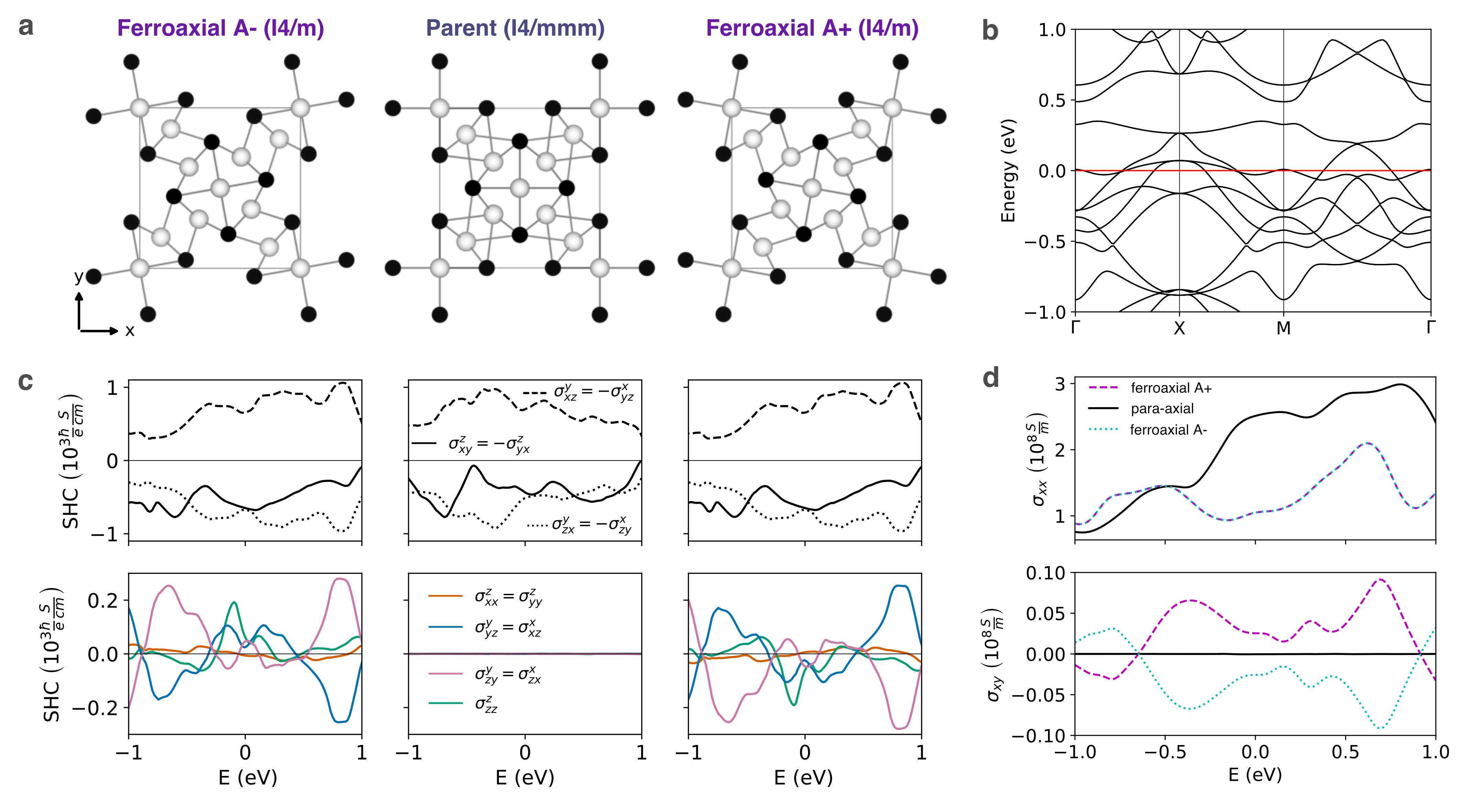}
    \caption{\textbf{First-principles calculations of ferroaxial metal Ta$_5$As$_4$.} (a) Top view of the crystal structure of Ta$_5$As$_4$ in para-axial (I4/mmm) and ferroaxial  (I4/m) phase, with Ta in white and As in black. A structural rotation appears in the ferroaxial phase, with opposite orientation in the two domains. (b) Band structure of Ta$_5$As$_4$ in the ferroaxial phase. (c) Spin Hall conductivity for the parent and ferroaxial phase of Ta$_5$As$_4$. The black solid, dotted and dashed curves on the first row are the conventional components. The colored solid curves on the second row are the unconventional components only present in the ferroaxial phase. (d) $xx$ and $xy$ component of conductivity, assuming a constant relaxation time of $10^{-12}$ s, with applied strain $\epsilon_{xx}=-1\%$. They are related to the conventional and unconventional piezoresistivity $\pi_{xxxx}$ and $\pi_{xyxx}$ respectively.}
    \label{fig:TaAs}
\end{figure*}

The ferroaxial materials discovered to date are mostly insulating, with canonical examples of Cu$_3$Nb$_2$O$_8$ \cite{Johnson_2011}, CaMn$_7$O$_{12}$ \cite{Johnson_2012} and RbFe(MoO$_4$)$_2$ \cite{Hearmon_2012}, while ferroaxial metals remained essentially unexplored until recently \cite{Oppeneer2026}. Here, we use the unconventional SHC components as descriptors to identify candidate ferroaxial materials among the nonmagnetic metals included in the spin Hall conductivity database of Ref. \cite{Zhang_2021}. The search identifies several families of metallic ferroaxial materials, as listed in the SM (Sec. VI). Here, we analyze representative compounds from two of these families in detail: the A$_5$B$_4$ intermetallics, where A is a transition metal and B is a pnictogen or chalcogen, and the quasi-one-dimensional A$_x$Nb$_6$X$_8$ compounds, where X is a chalcogen.

Experimentally synthesized compounds in A$_5$B$_4$ family include \{Ti, V, Nb\}$_5$Se$_4$, \{Ti, V, Zr, Nb, Hf\}$_5$Te$_4$, \{Mo, Ta\}$_5$As$_4$, \{V, Nb, Ta\}$_5$Sb$_4$ \cite{Jain_2013,Horton_2025}. We focus on Ta$_5$As$_4$ which crystallizes in space group I4/m (87) with a tetragonal lattice (Fig. \ref{fig:TaAs}a). Since a para-axial parent structure is not found experimentally, we construct a hypothetical one in space group I4/mmm (139) based on the Wyckoff positions in the ferroaxial phase. We perform first-principles calculations on both the ferroaxial and para-axial phase. After fully relaxing the structures, wavefunctions are computed and used to generate Hamiltonians in a pseudo-atomic orbital (PAO) basis for further calculations of SHC \cite{Buongiorno_Nardelli_2018, Cerasoli_2021}. Our relaxed ferroaxial structure exhibits a rotation angle of 10.2$^\circ$, close to the experimental values of 10.3$^\circ$ and 11.9$^\circ$ \cite{Dronskowski_1992, Rundqvist_1969}. We also computed conductivity in the strained lattice with relaxed internal coordinates as a way to obtain piezoresistivity. Computational details and comparison with experimental structures can be found in SM (Sec. VII).

We compare the responses of the para-axial and ferroaxial structures to verify the relation between ferroaxial order and unconventional tensor components. In the middle column of Fig. \ref{fig:TaAs}c, only the conventional SHC components $\sigma^k_{ij}$ ($i\neq j\neq k$) are present in the para-axial phase. In the left and right columns, corresponding to the ferroaxial phase, the unconventional components additionally appear, in agreement with our theoretical prediction. Comparing the SHC of opposite ferroaxial domains, we find that the conventional components remain unchanged, whereas the unconventional ones reverse sign, demonstrating the coupling between the unconventional spin Hall effect and ferroaxial order. 

The same ferroaxial control holds for other responses, including those described by rank-4 tensors. As an example, we calculate the piezoresistive response by applying a strain $\epsilon_{xx}=-1\%$, which produces an off-diagonal conductivity $\sigma_{xy}$, corresponding to the unconventional piezoresistivity coefficient $\pi_{xyxx}$, in addition to the conventional one $\pi_{xxxx}$ (see Box 1). Figure \ref{fig:TaAs}d demonstrates that unconventional piezoresistivity changes sign upon reversal of ferroaxial order. Notably, this provides a metallic realization of the ferroaxial piezoresistive response without requiring carrier doping \cite{Day_Roberts_2025}. 

A second family identified by our search comprises the quasi-one-dimensional metallic compounds A$_x$Nb$_6$X$_8$ in space group P6$_3$/m (176), where A = Na, K, Rb, Ca, Cu, Ag, Zn, Cd, Pb, Bi, and X = S, Se, Te and $0\leq x\leq2$ \cite{Huan_1987,Huan_1987_Se}. Our calculations for Nb$_6$Te$_8$ find a large unconventional SHC. The complete results and computational details are provided in SM (Sec. VII). Since the A atom is intercalated between the skeleton formed by Nb and X, we expect other A$_x$Nb$_6$Te$_8$ to exhibit SHC of similar magnitude. Additionally, $x$ may be varied to tune the Fermi level and potentially achieve larger SHC, which makes this family promising for spintronics applications and realizing other unconventional responses. 

\section*{Discussion}
Our results establish that ferroaxial order can be read out through a broad range of physical responses, each directly coupled to the ferroaxial state, providing a variety of probes through measurable electrical, optical, and mechanical signals. The absence of stray and depolarizing fields and the strongly nonvolatile character of ferroaxiality further favor its use in miniaturized electronic devices, where different material responses can be selected according to the specific requirements of a device architecture.

Beyond providing readout mechanisms, our results also reveal ferroaxiality as a playground for unconventional physical phenomena that are difficult to realize in conventional crystals. A prominent example is the collinear spin Hall effect, in which an electric current generates a transverse spin current with spin polarization parallel to the spin current \cite{Roy_2022}. Such spin currents are of particular interest for spin-orbit torque applications, where they can generate spin accumulation capable of switching a perpendicularly magnetized layer. Ferroaxiality provides an additional nonvolatile degree of control of such a response, extending ferroic control of electronic transport beyond conventional polarization and magnetization.

In addition, unlike approaches based on engineered crystal structures, interfaces, or superlattices, ferroaxiality provides unconventional responses intrinsically within the crystal. This may be particularly relevant for higher-order transport phenomena. Although inversion symmetry forbids the lowest-order nonlinear conductivity and magnetoconductivity in ferroaxial systems, higher-order responses can be allowed and provide access to quantum geometric effects beyond conventional transport \cite{Suarez2025}.

Our theoretical approach can be extended beyond ferroaxial order to identify which components of a response tensor transform according to a given ferroic order parameter. It could be applied, for example, to magnetic toroidal order. Moreover, in magnetic ferroaxial materials, the magnetic order breaks all vertical mirror symmetries \cite{Hayami_2022_M,Yang_2025}. In some cases, $\mathcal{T}$ symmetry is also broken, resulting in black-white or colorless magnetic groups. Then, additional responses that are not permitted in grey groups can arise, such as the magnetic ($\mathcal{T}$-odd) spin Hall effect, or recently suggested nonlinear Hall effect \cite{Agterberg2026}. Extending the analysis to other ferroic orders and including magnetic symmetry could reveal further classes of unconventional phenomena.

The growing range of materials and potential to fabricate single-domain ferroaxial crystals opens several directions to explore \cite{Fang_2023}. Our analysis provides a way to expand materials portfolio by using unconventional responses as descriptors for materials discovery. We demonstrated this approach by screening for ferroaxial metals using the unconventional spin Hall conductivity and identified representative 3D candidates that host such responses. This strategy can also be applied to 2D systems. For example, a recent high-throughput study find 17 2D metals of low symmetry with unconventional spin Hall components \cite{Zhou_2025}. Among them, Y$_2$C$_2$I$_2$ in space group C2/m (12) presents a high $\sigma^x_{xy}$ at the Fermi energy. Furthermore, ferroaxial density waves with an electronic origin in layered compounds 1T-TaS$_2$ \cite{Liu_2023} and RTe$_3$ (R = La, Gd) \cite{Wang_2022,Singh_2025} have been observed, whose unconventional responses could also be explored.

Overall, our results reveal ferroaxial order as an origin of a broad class of unconventional material responses. The symmetry connection established here links the ferroaxial order parameter to specific tensor components across different physical phenomena, providing a common basis for understanding how these responses emerge and evolve upon switching between ferroaxial domains. More broadly, ferroic orders provide a powerful route to nonvolatile control of electronic properties. Ferroaxiality adds a distinct member to this family - whereas ferromagnetism and ferroelectricity manifest mostly through magnetization and polarization and conventional responses, ferroaxiality manifests primarily through unconventional optical, charge, and thermal transport, and electromechanical phenomena. Ferroaxiality therefore substantially expands the range of physical phenomena that can be controlled through a nonvolatile ferroic order.

\section*{Data availability}
The data will be publicly available at DataverseNL after the manuscript is accepted. 

\section*{Code availability}
PAOFLOW is freely available at \url{https://github.com/marcobn/PAOFLOW} under the GNU General Public License v3 and can be installed from the PyPi repository (\texttt{pip install PAOFLOW}). Documentation, examples and tutorials can be found at \url{https://paoflow.org}.

\section*{Acknowledgments}
We thank Sergio Alvarruiz and Chao Chen Ye for fruitful discussions. The calculations were carried out on the Dutch national e-infrastructure with the support of SURF Cooperative (EINF-17379) and on the Hábrók high performance computing cluster of the University of Groningen.

\section*{Author contributions}
Z.H. performed all the group theory analysis and first-principles calculations, and wrote the initial manuscript draft. G.K. and J.S. supervised the project. 

\section*{Funding}
This research was funded by the EU through the ERC Consolidator Grant FERRERO (No. 101170522). 

\section*{Competing interests}
The authors declare no competing interests. 

\bibliography{main}

\end{document}

%% file: box_1/boxmacros.tex
\ifdefined\boxmacrosloaded   \fi
\let\boxmacrosloaded\relax

\usepackage{newfloat}
\DeclareFloatingEnvironment[fileext=lob,name=BOX,placement=tbp]{boxfloat}

\newlength{\effrowskip} 
\newlength{\boxsecskip} 
\newlength{\boxtopskipL}
\newlength{\boxtopskipR}
\newlength{\effalignlift}

\newlength{\boxinner}   
\newlength{\boxhalf}    

\providecommand{\boxcaptiontext}{this is the caption}
\providecommand{\boxlabelname}{box:definitions}

\newsavebox{\effsavebox}
\newlength{\boxtopskip}   
\newif\ifboxfirsthead

\newenvironment{effectbox}[2]
  {\def\effcaption{#1}\def\efflabel{#2}%
   \begin{boxfloat*}%
   \setlength{\boxinner}{\dimexpr\textwidth-2\fboxsep-2\fboxrule\relax}%
   \setlength{\boxhalf}{\dimexpr0.5\boxinner-1em\relax}%
   \begin{lrbox}{\effsavebox}%
   \footnotesize
   \begin{minipage}{\boxinner}%
   \leftskip=0pt \rightskip=0pt \parfillskip=0pt plus 1fil\relax}
  {\end{minipage}%
   \end{lrbox}%
   \centering
   \fbox{\usebox{\effsavebox}}%
   \caption{\effcaption}\label{\efflabel}%
   \end{boxfloat*}}

\newenvironment{boxcol}[1][\boxtopskipL]
  {\begin{minipage}[t]{\boxhalf}\setlength{\boxtopskip}{#1}\boxfirstheadtrue}
  {\end{minipage}}

\newcommand{\boxhead}[1]{%
  \par
  \ifboxfirsthead\boxfirstheadfalse\vspace*{\boxtopskip}%
  \else\addvspace{\boxsecskip}\fi
  \centerline{\textbf{#1}}\par\smallskip}

\newenvironment{effalign}
  {\setlength{\abovedisplayskip}{-\effalignlift}%
   \setlength{\abovedisplayshortskip}{-\effalignlift}%
   \setlength{\belowdisplayskip}{0pt}\setlength{\belowdisplayshortskip}{0pt}%
   \csname align*\endcsname}
  {\csname endalign*\endcsname}

\newenvironment{efflist}
  {\par\centering\begin{tabular}{@{}r@{\;}l@{\qquad}l@{}}}
  {\end{tabular}\par}
\newcommand{\eff}[3]{$#1$ & $#2$ & #3\\}
\newcommand{\effsep}[3]{\noalign{\vspace{\effrowskip}}$#1$ & $#2$ & #3\\}

\newcommand{\boxdivider}{\par\medskip\hrule\medskip}

\newenvironment{notation}{\par\centering\textbf{} \ignorespaces}{\par}
\newcommand{\nt}[2]{\mbox{$#1$ = #2};\ }
\newcommand{\ntlast}[2]{\mbox{$#1$ = #2}.}

%% file: box_1/box.tex

\begin{effectbox}{\boxcaptiontext}{\boxlabelname}

\begin{boxcol}[\boxtopskipL]
\input{box_1/sec_linoptical}
\input{box_1/sec_lintransport}
\end{boxcol}\hfill%
\begin{boxcol}[\boxtopskipR]
\input{box_1/sec_equilibrium}
\input{box_1/sec_nonlinear}
\end{boxcol}

\input{box_1/sec_notation}
\end{effectbox}

%% file: box_1/sec_linoptical.tex
\boxhead{Linear optical effects}
\begin{efflist}
\eff{\Delta\beta_{ij} =}{F_{ijk}H_k}{Faraday effect}
\eff{+}{R_{ijk\ell}E_kE_\ell}{Kerr effect}
\eff{+}{C_{ijk\ell}H_kH_\ell}{Cotton-Mouton effect}
\eff{+}{p_{ijk\ell}\varepsilon_{k\ell}}{Elasto-optical effect}
\effsep{\Delta g_{ij} =}{\gamma_{ijk}E_k}{Electrogyration effect}
\end{efflist}

%% file: box_1/sec_lintransport.tex
\boxhead{Linear transport effects}
\begin{efflist}
\eff{E_i =}{\rho_{ij}J_j}{Electric resistivity}
\eff{+}{\beta_{ij}\nabla_jT}{Seebeck effect}
\eff{+}{R_{ijk}J_jH_k}{Hall effect}
\eff{+}{N_{ijk}\nabla_jTH_k}{Nernst effect}
\eff{+}{T_{ijk\ell}J_jH_kH_\ell}{Magnetic resistance}
\eff{+}{\alpha_{ijk\ell}\nabla_jTH_kH_\ell}{Magneto-Seebeck effect}
\eff{+}{\sigma_{ij}^k\mathcal{J}_j^k}{Inverse Spin Hall effect}
\effsep{q_i =}{T\pi_{ij}J_j}{Peltier effect}
\eff{+}{Q_{ijk}\nabla_jTH_k}{Thermal Hall effect}
\eff{+}{N_{ijk}J_jH_k}{Ettinghausen effect}
\eff{+}{P_{ijk\ell}J_jH_kH_\ell}{Magneto-Peltier effect}
\effsep{\Delta\rho_{ij} =}{\pi_{ijk\ell}\sigma_{k\ell}}{Piezoresistivity}
\end{efflist}

%% file: box_1/sec_equilibrium.tex
\boxhead{Equilibrium effects}
\begin{efflist}
\eff{P_i =}{\chi_{ij}^{e,\mathrm{eq}}E_j}{Dielectric susceptibility $\chi_1^{e,\mathrm{eq}}$}
\eff{+}{\chi^{e,\mathrm{eq}}_{ijk\ell}E_jE_kE_\ell+\ldots}{and $\chi_3^{e,\mathrm{eq}}\ldots$}
\eff{+}{\mu_{ijk\ell}\nabla_j\varepsilon_{k\ell}}{Flexoelectricity}
\effsep{\varepsilon_{ij} =}{S_{ijk\ell}\sigma_{k\ell}}{Elastic compliance}
\eff{+}{\gamma_{ijk\ell}E_kE_\ell}{Electrostriction}
\eff{+}{N_{ijk\ell}H_kH_\ell}{Magnetostriction}
\effsep{M_i =}{\chi_{ij}^{m,\mathrm{eq}}H_j}{Magnetic susceptibility $\chi_1^{m,\mathrm{eq}}$}
\eff{+}{\chi^{m,\mathrm{eq}}_{ijk\ell}H_jH_kH_\ell+\ldots}{and $\chi_3^{m,\mathrm{eq}}\ldots$}
\end{efflist}

%% file: box_1/sec_nonlinear.tex
\boxhead{Nonlinear optical and transport effects}
\begin{effalign}
    P_i(\omega_4)={}&
    \chi_{ijk\ell}(\omega_4;\omega_3,\omega_2,\omega_1)
    E_j(\omega_3)E_k(\omega_2)E_\ell(\omega_1),\\
    & \text{General third-order susceptibility, $\chi_3$} \\
    &\begin{cases}
        \omega_1=\omega_2=\omega_3=\omega,\ \omega_4=3\omega
        & \text{THG},~\chi_3^{\mathrm{THG}}\\
        \omega_1=\omega_2=\omega,\ \omega_3=0,\ \omega_4=2\omega
        & \text{EFISH},~\chi_3^{\mathrm{EFISH}}\\
        \omega_3=-\omega_1,\ \omega_4=\omega_2
        & \text{FWM},~\chi_3^{\mathrm{FWM}}\\
        \omega_1=\omega_2=-\omega_3=\omega_4=\omega
        & \text{DFWM},~\chi_3^{\mathrm{DFWM}}
    \end{cases}\\[\effrowskip]
    J_i(\omega)=&-i\omega P_i(\omega) \rightarrow~ \text{Nonlinear transport}~\sigma_3=-i\omega\chi_3 \\[\effrowskip]
    M_i(\omega_3)=&
    \chi^{m}_{ijk}(\omega_3;\omega_2,\omega_1)
    E_j(\omega_2)E_k(\omega_1)\\
    &\text{General second-order susceptibility by magnetic dipole, $\chi_2^m$}\\
    &\text{}
\end{effalign}

%% file: box_1/sec_notation.tex
\boxdivider
\begin{notation}
\nt{T}{temperature}
\nt{E}{electric field}
\nt{P}{electric polarization}
\nt{H}{magnetic field}
\nt{M}{magnetization}
\nt{J}{charge current}
\nt{\mathcal{J}}{spin current}
\nt{q}{heat current}
\nt{\sigma}{stress}
\nt{\varepsilon}{strain}
\nt{\Delta\beta}{change in dielectric impermeability tensor}
\nt{\Delta g}{change in gyration tensor}
\ntlast{\Delta\rho}{change in resistivity tensor}
\end{notation}